\documentclass{article}

\usepackage{amssymb}
\usepackage{mathtext}
\usepackage{color}
\usepackage{graphicx}
\usepackage{epsf}
\usepackage{quant}
\title{Classification of two-particle quantum channels of information transfer}
\newcommand{\bit}[1]{\left\lfloor #1\right\rfloor}

\author{Constantin V. Usenko}
\date{}
\begin{document}
\maketitle
\tableofcontents
\begin{abstract}
Classification of states of two-particle quantum channels of information transfer is built on the basis of unreducible  representations of qubit state space group of symmetry and properties of density matrix spectrum. It is shown that the reason of state disentanglement can be in degeneration of non-zero density matrix eigenvalues. Among the states with non-degenerate density matrix disentangled states form two-dimensional surface of special states.
\end{abstract}

\section{Introduction}

 Fundamental difference of propagation and interaction of quantum information carriers laws from classical ones leads to new in principle information possibilities of quantum channels of information transfer. Main distinction for quantum carriers is in existence of states of parallel quantum channels of information transfer with nonremovable correlations between channel sections - the entangled states. Considerable compared to classical parallel channels complexity of information transfer quantum channels state space leads to necessity to use, to measure the state completely, complicated experimental methods, like quantum tomography  \cite{Manko}, and to describe state set -- overfull basis of state space \cite{wootters89, MUBDWF}. Complete study of properties of entangled states is possible under condition of construction of complete classification of entangled states, this is to rest upon properties of symmetry of the component parts of parallel quantum channel of information transfer state properties -- qubits.
Some success in use of qubit state space symmetry for description of entangled states is just achieved \cite{symbqs,symiso,symws,Rotation,schliemann}, though the problems on indispensable symmetry of composite quantum systems, and on completeness of state classification by unreducible representation of qubit state space symmetry remain unresolved. Suitable solution of these two problems is main point of this manuscript.

Classical information theory is to a considerable degree based on statement that information carrier with finite number of possible states can be represented by means of a set of two-state information carriers -- bits. Basis for such a point of view is in the fact that arbitrary measurement (and measurement is just what gives information) has as its result each time one of finite set of possible values only. Really, even measurement of values we usually consider to be continuous takes places with some error and the results differing less than by error value are treated as same. Finite set of possible values can in its turn be numbered. Representation of the value number in binary format is given by a set zero/one, this is bitwise representation of information carrier state. 

Common peculiarity of all the implementations of bit is in the fact that its states belong to two-element space $B_2$.  

As quantum analog of bit one can consider two-quantum channel of information transfer, with two possible states, this is given the name "`qubit"'. Examples of qubit are two-level atom, electron with two possible spin directions, photon with two possible directions of polarization. 

Common peculiarity of qubits is specific character of state measurement statistics. For each pure state of arbitrary quantum system there exists just one detector for which probability of registration of system as whole is equal to one. For qubit in given pure state there exists as well just one detector for which probability of registration of qubit in that state is equal to zero. The simplest example is measurement of linear polarization of photons by means of Nicol prism: for photon in state with polarization parallel to the plane of polarization of prism probability of registration in normal ray is equal to one, and in abnormal one - zero, and other possibilities for photon registration Nicol prism does not provide.

Information that is to be transferred is usually coded by a set of bits, not by one bit only. Those bits can be transferred by one channel in sequence, or  by several channels simultaneously; in both cases it is considered that mutual effect of separate bits is not present, or can be compensated by technical means. Because of that, basic object of study in classical information theory is just the bit, and main problems are in determination of characteristics of the channel restricting rate of information transfer, methods for compensation of loss of information, methods for channel reliability evaluation. Channel is an arbitrary device for information processing, even one that stores information during some time only.

Sequential multibit quantum channels are distinguished from classical ones by properties of each separate qubit only and there is no need in specific study for those. Parallel channels differ from classical ones in principle by possibility of existence of non-removable correlations between separate subchannels. In this work the problem on classification of states of two-particle information transfer quantum channels is studied. Effectiveness of the classification methods proposed is shown on two important examples - paraqubit that is common state of pair of two-state particles, and paraqutrit being composite state of qubit and qutrit.

\subsection*{Parallel channels for information transfer}

One of the first manifestations of quantum nature is in ability of electromagnetic field to be emitted/absorbed in integral parts only -- in quanta of electromagnetic field, or photons. Classical information transfer channels perform transfer in separate portions -- bits as well, though that is achieved through special technical ideas; at the same time development of source of separate photons is today an important scientific, if not yet technical, problem \cite{onesrc}. 

Classical channels for information transfer are sometimes parallel because of transfer speedup only. Parallel channel is just like sequential one, though instead of the instant of transfer sequential bit is identified with the channel number. Information transferred by system of parallel channels is function of states of separate channels. 

Quantum parallel channels, in spite of classical ones, can have nonremovable, quantum by nature, correlations, since the set of channel system states is larger than the product of the sets of sub-channel states. At first sight, the last statement comes into contradiction with linearity of Hilbert spaces vectors of which represent the states of arbitrary quantum system. In fact the states, at least the pure ones, are represented not by arbitrary vectors of state space $\cat{\psi}$ though by rays $\left\{W\cat{\psi}:\forall W\in C \& W\neq0\right\}$ --  all the vectors differing by multiplier correspond to one state. The number of independent (real) parameters of arbitrary pure state in  $N$-dimensional space of one subsystem is $2N-2$, in $M$-dimensional space of the second subsystem -- $2M-2$, and in $N*M$-dimensional space of composite system is  $2N*M-2$ and exceeds the total number of subsystem parameters by value $2N+2M-4$, that is above zero for all nontrivial compositions.  
For instance, for paraqubit with four-dimensional state space there exist six arbitrary parameters of pure state, whereas each qubit has two parameters only. Like that, pair of qutrits has six-dimensional state space, there exist ten arbitrary parameters of pure state, whereas combination of qubits with qutrits has six parameters only (four -- qutrit, and two -- qubit).

As the result, not each pure state of composite system is composition of pure states of its sub-systems, some, and rather large part of pure states is formed by mixed states of sub-systems.

Mixed states of channel, quantum or classical ones, differ from pure ones by the fact that those are prepared, with given frequency, in different pure states. There exists principal possibility to perform control of the fact in which specific pure state the channel has been prepared in sequential transfer event. At the same time in the sequence of transfer events in which a channel has been prepared for the first  $p_1 N$  times in first state, then $p_2 N$ times in the second one, and so on, statistics of measurements does not differ from same series in which, occasionally or with prediction, the channel has been prepared in different states in another, more complicated, sequence, though in such way that total number of cases of the first state is equal to the same value $p_1 N$, of the second one -- $p_2 N$.  Thus information coded in distribution of mixed channel state by pure states is placed to channel in the process of state formation, and difference of channel state from planned one can be just a result of imperfection of devices preparing the channel for transfer. 

Pure states of the channel are prepared in same way in each event of information transfer, and statistic deflections of results of measurements of channel state observable characteristics are manifestation of quantum uncertainty, and make evidence of mismatch of measuring device and the source. In the case of total match pure states are registered with respective detector with probability one, and are not registered with any one other -- statistic distribution of channel state registration is not present. Such match of the source of quantum system states and measuring device is known as nondestructive measurement.

 In parallel information transfer quantum channel set of measuring devices for subchannel states able to provide nondestructive measurements exists not always. Each of pure states of parallel channel that can not be given by combination of independent states of subchannels can not be measured without state reduction with devices that register the state of each subchannel independently. At interaction of subchannel particles with state analyzers of separate channel sections mixed subchannel states are formed, for those statistic uncertainty of measurement result is always specific. So, statistic uncertainty comes to existence in the process of measurement of component parts of parallel quantum information transfer channel.  
  
  Information emerging in each event of transfer of pure state of parallel quantum channel in the process of measurement of state of one part of channel can not be, usually, information on the state of channel as whole, thus it is to be related to the state of other parts of the channel, and this is possible in the case of correlation of results of measurements for parallel channel parts only. Due to correlation properties with specific quantum origin we come to idea of entanglement of states.

  Correlation between channel parts is present as well in the case of channel being prepared at some part of transfer events with one part in first state, and the second in first as well, the second part of transfer events -- with parts in  second states, and so on. Such state of parallel quantum channel is mixed. Information is put to such channel in the process of channel state preparing, instead of being created in the process of measurement. Correlations due to entanglement of states qualitatively differ from classical ones, Bell inequalities make evidence of it.

  Study of specific properties of quantum parallel information transfer channels needs use of specific measure of those properties. Physical characteristic of unremovable statistic uncertainty is von Neumann entropy. It coincides with Shannon entropy in own basis of density matrix and is equal to zero for pure states. Shannon entropy, as independent from physical realization measure of uncertainty of probability distribution, has specific properties as to division of statistic system to subsystems. So, for a system consisting of independent subsystems Shannon entropy is equal to the sum of subsystem entropies, while in opposite case of completely correlated system Shannon entropy for each subsystem is equal to entropy of system as a whole.

   Entangled pure states are distinguished by non-zero Neumann entropy of subsystems, and at the same time with zero one -- of the system. Excess of entropy coming to existence because of system division into subsystems has got the name of relative entropy of entanglement \cite{VP}). Later on some problems on effectiveness of entanglement evaluation by this value have been shown \cite{RHMH} and other measures  \cite{Wootters}  and criteria \cite{PHPLA,PHPRL} of entanglement have been developed.

\section{Qubit states}
Description of physical properties of qubit is given by its states. Pure states are given by normalized vectors of two-dimensional Hilbert state $\mathcal{H}=C^2$. Analogs of classical states are specific states of quantum system $\cat{0}$ and $\cat{1}$, those can be chosen as basis of qubit state space. 
\subsection{Pure states}
Arbitrary qubit pure state is characterised by state vector 
\begin{equation}
	\cat{\psi}=c_0\cat{0}+c_1\cat{1}.
\end{equation}
State vector is linear combination of basis vectors, and thus it is marked by two complex numbers $c_0$ and $c_1$ -- coefficients of state representation in given basis. Those numbers are not completely independent since all the vectors differing by nonzero multiplier, common for both coefficients, only,  $\tilde{c}_0=Cc_0$, $\tilde{c}_1=Cc_1$, give one physical state. Most of all as representative of state one of normalized vectors $\abs{c_0}^2+\abs{c_1}^2=1$ is chosen, then there is left possibility to multiply coefficients by arbitrary phase multiplier $\tilde{c}_0=e^{i\psi}c_0$, $\tilde{c}_1=e^{i\psi}c_1$ only.

Arbitrary pure state 
\begin{equation}
	c_0=e^{i\psi}\cos\theta/2;\ c_1=e^{i\psi+i\phi/2}\sin\theta/2,
\end{equation}
is completely determined by two real parameters $\phi$, $\theta$, set of which consides with the set of points of sphere named as Poincare sphere (quantum optics) or Blokh sphere (solid state physics). 

Quantum analog of classical bit state pair is each pair of opposite in diameter points of Poincare sphere. Each of those pairs consists of orthogonal to each other states, and sets vector pair
\begin{equation}
\begin{array}{ll}
	\cat{0}=&\cat{\theta,\phi};\\
	\cat{1}=&\cat{\pi-\theta,\left(\phi+\pi\right) \mathop{\rm mod} 2\pi};
	\end{array}
\end{equation}
forming the basis in the qubit state space. 

Instead of point pair of discrete space of classical channel states quantum channel gives for transfer all the set of opposite in diameter points of Poincare sphere. 

\subsection{Qubit density matrix}

Density matrix gives more universal method for state description.
 
For arbitrary pure state density matrix is projector to one-dimensional subspace of vectors collinear to state 
\begin{equation}
	\hat{\rho}\left(\psi,\theta\right)=\cat{\psi,\theta}\otimes\bra{\psi,\theta},
\end{equation}
and thus in given basis it has following representation
\begin{equation}
  \begin{array}{ll}
	\hat{\rho}\left(\theta,\phi\right)&=\left(1+\cos\theta\right)/2\cat{0}\bra{0}+
	\left(1-\cos\theta\right)/2\cat{1}\bra{1}\\
	&+
	\left(\sin\theta\right)/2\left(e^{-i\phi/2}
	\cat{0}\bra{1}+e^{i\phi/2}\cat{1}\bra{0}\right)
	\end{array}.
\end{equation}
Information transfer quantum channel that with given frequency $p$ gets from the source state $\cat{1}$, and with frequency $1-p$ -- state $\cat{0}$, has mixed qubit state that can be given as mix of pure state $\cat{0}$ with weight $1-2p$ and equilibrium with weight $2p$
\begin{equation}
	\hat{\rho}\left(p,\theta,\phi\right)=p\cat{0}\bra{0}+
	\left(1-p\right)\cat{1}\bra{1}
	=\left(2p\right)\frac{1}{2}\hat{1}+\left(1-2p\right)\hat{\rho}\left(\theta,\phi\right).
\end{equation}

\subsection{Measurement and reduction of states}
Correlation of states of multiparticle quantum information transfer channel is observed through measurement of states, thus complete classification of correlation properties of states rests on detailed analysis of peculiarities of few-particle composite state. 

Process of measurement of state of arbitrary quantum system is accompanied by reduction of state to one of eigenstates of measuring device. Complete description of measurement process is based on separation of state and detection of results of analysis. 

In the process of analysis interaction of the particle being measured and analyzer takes place. Typical examples are interaction of polarized photon with Nicol prism, or electron with magnetic field. Result of such interaction is in change of direction of motion of photon depending on its polarization, or electron depending on spin orientation with respect to magnetic field. Important is as well the fact that photon or electron moves in given direction even in the case of polarization (or spin) before interaction with analyzer not corresponding to some of possible end states, i.e. in the case of particle state before analyzer being superposition of base states.

\subsubsection{Measurement of quantum particle}

 At the beginning of interaction density matrix of the system "`particle+analyzer"' is product of density matrices of independent subsystems $\hat{\rho}_f=\hat{\rho}_p\otimes\hat{\rho}_a$. 
  As the result of interaction it is transformed to mix (this transformation is not unitary!)
\begin{equation}\label{decomp}
	\hat{\rho}_f=\sum_{\forall k}{p_k\hat{\rho}_p^{\left\{k\right\}} \otimes\hat{\rho}_a^{\left\{k\right\}}}.
\end{equation}
This mix consists of products of density matrices of independent states of particle and analyzer  $\hat{\rho}_f^{\left\{k\right\}}=\hat{\rho}_p^{\left\{k\right\}}\otimes\hat{\rho}_a^{\left\{k\right\}}$. 
Weight of each product is determined by the portion of respective state of particle 
\begin{equation}
	p_k=\tr{\hat{\rho}_p^{\left\{k\right\}}\hat{\rho}_p},
\end{equation}
in its initial state. This very value is to be measured, as expected portion of $k$-th possible state in the state of particle being measured.

 It is considered that all the density matrices of substates $\hat{\rho}_p^{\left\{k\right\}}$ are orthogonal to each other 
\begin{equation}
	 \hat{\rho}_{p,a}^{\left\{k\right\}} \hat{\rho}_{p,a}^{\left\{m\right\}}=\hat{\rho}_{p,a}^{\left\{k\right\}}\delta_{k,m},
\end{equation}
  and form complete system 
\begin{equation}
	  \sum_{\forall k}{\hat{\rho}_{p,a}^{\left\{k\right\}}}=\hat{1}_{p,a}. 
\end{equation}
After particle interaction with analyzer detection of the results of interaction takes place. The simplest version of detection is absorption of particle with some detector of total set. In this variant it is supposed that each of substates of the particle  $\hat{\rho}_p^{\left\{k\right\}}$ can be absorbed (with probability 1) by one of detectors only, and can not be absorbed by any other detector, thus relative portion of counts of each of detectors gives evaluation of probability of respective substate.

More sophisticated is detection of the states of analyzer. Advantage of this variant is the fact that the particle is left after analyzer in well-known state, such measuring device is at the same time a generator of given states. Substantial disadvantage of such generator is in fact that the state is created with some probability smaller than one, thus not in each event of generation just the prescribed, specific needed state is created. 
\subsubsection{Qubit measurement}

In two-dimensional state space orthogonal projectors form pairs, for each given projector \[\hat{P}_0\left(\theta,\phi\right)=\cat{\theta,\phi}\bra{\theta,\phi}\]
there exists one projector orthogonal to the given one \[\hat{P}_1\left(\theta,\phi\right)=\cat{\pi-\theta,2\pi-\phi}\bra{\pi-\theta,2\pi-\phi}
=\hat{1}-\hat{P}_0\left(\theta,\phi\right)
.\]
As the result, density matrix of arbitrary state after interaction with analyzer is transformed to mix of eigenstate of analyzer and white noise
\[\hat{\rho}\stackrel{\theta,\phi}{\rightarrow}
\begin{array}{ll}
\left(1-p\right)\frac{1}{2}\hat{1}+p\hat{\rho}\left(\theta,\phi\right);& p=2\mess{\theta,\phi}{\hat{\rho}}-1\geq 0\\
\left(1-\abs{p}\right)\frac{1}{2}\hat{1}+\abs{p}\hat{\rho}\left(\pi-\theta,2\pi-\phi\right);& p=2\mess{\theta,\phi}{\hat{\rho}}-1\leq 0\\
\end{array}.
\]
Only mix of eigenstates of analyzer remains non changed after interaction with analyzer. That's why the channel in which the states are prepared and measured by means of same (or coordinated) analyzers does not differ from classical channel by its properties.

The channel which states are prepared in arbitrary basis qualitatively differs from classical one since measurement of each specific state takes place with loss of part of information on state.

Orthogonal decomposition of unit operator acting in the qubit state space consists of two projectors $\hat{P}_0=\cat{0}\otimes\bra{0}$ та $\hat{P}_1=\cat{1}\otimes\bra{1}$, that for each pure state set amplitudes $c_{0,1}$ of probabilities of registration of particle with respective detector $\hat{P}_{0,1}\cat{\psi}=c_{0,1}\cat{0,1}$. For mixed state $\hat{\rho}$ probabilities of particle registration with respective detector are given by projection of density matrix $\hat{P}_{0,1}\hat{\rho}\hat{P}_{0,1}=p_{0,1}\hat{P}_{0,1}$ to substate corresponding to the detector. For pure states, and for mixed ones as well, as the result of projecting part of state parameters is lost. For pure state one of two state parameters is lost -- the value of angle  $\phi$ on the Poincare sphere, whereas for the mixed state from three state parameters only one is left. 

\subsubsection{Quantum tomography}
Determination of all the three parameters of qubit state needs at least three independent measurements that use incompatible developments of unit operator. As an example of such measurements one can point out the following projector sets: $P_{z|0,1}=\frac{1}{2}\left(1\pm \sigma_z\right)$, $P_{x|0,1}=\frac{1}{2}\left(1\pm \sigma_x\right)$, 
$P_{y|0,1}=\frac{1}{2}\left(1\pm \sigma_y\right)$. 
Each of measurements make it possible to determine one of vector $\vec{r}$ components only that characterizes the Bloch matrix of the qubit.

Replacement of one representation of unit operator with another one is in the example given above similar to transition to measurement of Bloch vector  $\vec{r}$  protection to another axis, this is to be performed by "`turn"' of the measuring device in conventional phase space. That is why the method \cite{Manko} for quantum system measurement through performing measurements of needed number of noncommuting observables has got the name of quantum tomography.

\section{Entanglement of states}
Larger than finite power of state space of separate qubit in more complicated compared to two-state channels of information transfer is to make impossible reduction of state of complicated channel to simple combination of states of independent qubits. Along with that at least some (for instance, basis) states of complicated channel in the case of dimension of its state space being power of two can be given by direct product of needed number of qubit states. So, the states of complicated channel are to be divided to ones similar to classical states and ones that have no classical analog. The last ones are given the name of entangled states. In this section entangled composite states of multi-state system are considered.

\subsection{Multi-state systems}

Quantum information transfer channel with state space dimension $\dim \mathcal{H}=N$ larger than two is not to be called "`qubit"'. One can point out quantum properties and non-trivial space dimension together by means of term "`qunit"', combining quantum properties and $N$-dimensionality. 

Qunit arbitrary state can be given by coefficients of the state decomposition to series by vectors of orthonormalized basis $\left\{\cat{k};\ k=1\ldots N\right\}$
\begin{equation}
	\cat{\psi}=\sum_{k=1}^{N}{\psi_k\cat{k}};\ 
	\sum_{k=1}^{N}{\left|\psi_k\right|^2}=1.
\end{equation}
Because of normalization conditions and arbitrary phase multiplier set of values of qunit coefficients is equivalent to $2\left(N-1\right)$-dimensional sphere.
\subsubsection{Multiqubits (qudits)}
Specific variant of qunits are quantum information transfer channels with state space dimension equal to power of two $\dim \mathcal{H}=N=2^d$. 

Arbitrary state of classical information transfer channel with such dimension is split to direct product of separate bits $\bit{b_0,\ldots,b_{d-1}}=\bit{b_0}\otimes\ldots \otimes\bit{b_{d-1}}$, being states of separate one-bit classical information transfer channels. Just in that way classical channel transfers information with arbitrary complexity. 

Unlike classical channel, in quantum information transfer channel only some vectors can be given by products of basis vectors of qubits.
\begin{equation}
	\catt{b_0,\ldots,b_{d-1}}=\cat{b_0}\otimes\ldots\otimes\cat{b_{d-1}},
\end{equation}
though arbitrary state of qudit is more complicated than direct product of qubits. Really, qudit state is determined by $2\left(2^d-1\right)=2^{d+1}-2$ parameters, while the state of direct product of $d$ qubits -- $2^d$ parameters. Even the paraqubit ($d=2$) has 6 state parameters, whereas pair of separate qubits -- only 4. 

Existence of additional compared to qubit set parameters of multi-qubit state makes evidence of existence of additional properties of multi-qubit states that can not be reduced to properties of qubit set.  

\subsection{Paraqubit}
The simplest non-trivial qudit is a paraqubit -- system with four-dimension state space that one can make an attempt to represent by means of a pair of qubits.

Since pure qubit states are given by two real numbers, the number of parameters of qubit pair is not enough to bring those to correspondence with paraqubit pure state. The number of parameters becomes enough only under supposition of possibility of paraqubit pure state formation by specific combinations of mixed states of separate qubits. Those very pure states of paraqubits formed by mixed states have properties that can not be present in classical information transfer channels.

Let us use denotation $\catt{\#}$ for vectors in the state space of paraqubit, $\cat{\#}$, $\catr{\#}$ - the first and the second qubits. We suppose that there exists possibility to differ qubit states, for instance, by interaction with magnetic field (spin qubit) and electric one (spatial wave function in two-level atom -- spatial qubit).

For physical meaning of excess parameters of state of paraqubit one has to look for among additional variants of states differing from direct product of states of two qubits. 
\subsubsection{Paraqubit pure states}

Let us first consider paraqubit pure states. Those are given by state vector
\begin{equation}\label{para_pure}
	\catt{\psi}=\sum_{k,m=0,1}{\psi_{k,m}\cat{k}\otimes\catr{m}}.
\end{equation}
Four complex coefficients of this sum (eight real numbers) satisfy the normalization condition  $\sum_{k,m=0,1}{\abs{\psi_{k,m}}^2}=1$ and are defined with accuracy up to common phase multiplier, that's why there are 6 different real numbers setting the paraqubit state. Set of those numbers is 6-dimensional sphere.

State of first/second qubit is obtained by averaging of paraqubit density matrix by possible states of second/first qubit
\begin{equation}\label{para_to_one}
\hat{\rho}_{f}=\sum_{mm=0,1}{\brar{mm}\hat{\rho}_{para}\catr{mm}}=\sum_{m,k,kk=0,1}{\psi^*_{k,m}\psi_{kk,m}\cat{kk}\bra{k}};
\end{equation}
\[	\hat{\rho}_{s}=\sum_{kk=0,1}{\bra{kk}\hat{\rho}_{para}\cat{kk}}=\sum_{k,m,mm=0,1}{\psi^*_{k,m}\psi_{k,mm}\catr{mm}\brar{m}}.
\]

Unitary transformations of state spaces basis of the first and the second qubits make each of those density matrices diagonal. If one of eigenvalues of density matrix for the first/second qubit is equal to zero (the second one to one, respectively) qubit states are as well pure ones, and in some basis set paraqubit state is direct product of qubit states. Choosing respective qubit states as   $\cat{0,1}$ and $\catr{0,1}$, we have
\begin{equation}\label{pure_00}
	\catt{0,0}=\cat{0}\otimes\catr{0};\ \catt{0,1}=\cat{0}\otimes\catr{1};\ 
	\catt{1,0}=\cat{1}\otimes\catr{0};\ \catt{1,1}=\cat{1}\otimes\catr{1}.
\end{equation}
Such states correspond to classical parabit decomposition into two bits.

In the case of density matrices of first/second qubit eigenvalues differing from zero and one qubit states are not pure. One of examples of such state is EPR-state
\begin{equation}
	\catt{EPR}=\frac{1}{\sqrt{2}}
	\big(e^{i\phi/2}\cat{0}\otimes\catr{1}+e^{-i\phi/2}\cat{1}\otimes\catr{0}\big).
\end{equation}
Density matrices of the first and the second bit in this state are same and correspond to probability distribution of possible states
\begin{equation}
	\hat{\rho}_{f}=\frac{1}{2}\big(\cat{0}\bra{0}+\cat{1}\bra{1}\big);\ 
	\hat{\rho}_{s}=\frac{1}{2}\big(\catr{0}\brar{0}+\catr{1}\brar{1}\big). 
\end{equation}
Such state of paraqubit is completely non-classical since it has non-trivial properties not from standpoint of classical information theory only, but from standpoint of physics as well. Since EPR-state of paraqubit is pure its entropy is equal to zero, while entropy of each of bits is equal to one. Excess of entropy of subsystems over entropy of the system is an important pequliarity of nonclassical states that are given the name of entangled states.

\subsubsection{Mixed states of paraqubit}

Arbitrary mixed state of paraqubit is defined by its density matrix
\begin{equation}\label{para_mix}
	\hat{\rho}_{para}=\sum_{m,mm=1\ldots 4}{\rho_{m,mm}\catt{m}\brat{mm}},
\end{equation}
that has, being an hermitian matrix, 16 real numbers as independent parameters, and with account of normalization ($\sum_{m=1\ldots 4}{\rho_{m,m}}=1$) -- 15 independent real parameters. 
Through choosing as basis states direct products of qubits \eref{pure_00}, one has representation of density matrix
\begin{equation}\label{para_mix00}
	\hat{\rho}_{para}=\sum_{m,mm;k,kk=0,1}{\rho_{m,k;mm,kk}\catt{m,k}\brat{mm,kk}},
	\end{equation}
from which it is easy to obtain expressions for qubit density matrices
\[
	\hat{\rho}_{f}=\sum_{m,mm;k=0,1}{\rho_{m,k;mm,k}\cat{m}\bra{mm}},
\]
\[
	\hat{\rho}_{s}=\sum_{m;k,kk=0,1}{\rho_{m,k;m,kk}\catr{k}\brar{kk}}.
\]
With account of normalization and hermicity of density matrices one can write the relations between paraqubit and its component parts density matrices coefficients
\begin{equation}
	\begin{array}{ll}
	\rho^{\left\{f\right\}}_{00}=\rho_{00,00}+\rho_{01,01}&
	\rho^{\left\{f\right\}}_{01}=\rho_{00,10}+\rho_{01,11}\\
	\rho^{\left\{s\right\}}_{00}=\rho_{00,00}+\rho_{10,10}&
	\rho^{\left\{s\right\}}_{01}=\rho_{00,01}+\rho_{10,11}\\
\end{array}.
\end{equation}
Here it is taken to account that qubit is determined by one of two real diagonal coefficients of density matrix, for instance, $\rho^{\left\{f,s\right\}}_{00}$, and one of two complex non-diagonal coefficients $\rho^{\left\{f,s\right\}}_{01}$.

These equations are not enough to determine by coefficients $\rho^{\left\{f,s\right\}}_{00}$,  $\rho^{\left\{f,s\right\}}_{01}$ of both qubit state matrices all the coefficients $\rho_{m,k,mm,kk}$ of paraqubit state matrix, thus even complete study of separate components of paraqubit does not make it possible to restore its state, one has additionally analyze methods for quantum channel state measurements.
\section{Determination of state}
Determination of state of quantum system is one of kinds of measurement. Separate measurement can have as its purpose determination of average value of given observable, uncertainty of that value, or even probability distribution for observation of each possible value. Determination of quantum particle state is aimed at obtaining all density matrix of system components, and that's why it has to include the set of probability distributions for different non-compatible observables measurements.

Analysis of the problem of state determination is rational to begin from detailed description of state space of multi-state system. Quantum theory usually allows the observables to take not only discrete, but continuous spectrum as well, with supposing only that state space is Hilbert one, thus it has discrete basis. On the other part, classical information theory is in general restricted by consideration of information transfer channels characterized by state space with discrete number of points. Quantum generalization of such classical channels are quantum systems with finite basis of state space \cite{probab1,probab2}, where each basis vector corresponds to one classical state of channel, and due to quantum properties existence of superposition channel states becomes possible. 

\subsection{Probabilities and basis}
Set of sub-states \eref{decomp}, to which analyzer projects the state being measured is necessarily complete (otherwise part of states is not observed at all). It is desirable as well for that state to be orthogonal for independence of events, and for space of each of substates to be one-dimensional for completeness of measurements. Under such conditions each of substates is pure and is determined by state vector   $\cat{k}$. 
  Density matrix of such state 
\begin{equation}\label{det_basis}
	\hat{\rho}^{\left\{k\right\}}=\roa{k},
\end{equation}
has properties needed for projector 
\[\roa{k}\roa{kk}=\roa{k}\delta_{k,kk};\ \sum_{k=1\ldots N}{\roa{k}} =\hat{1},\]
and the vectors $\left\{\cat{k}:\ k=1\ldots N\right\}$ form the basis of state space.

Results of measurements are probabilities of registration of each of basis states
\begin{equation}
	p_k=\mess{k}{\hat{\rho}},
\end{equation}
or values of $N-1$ moments of some observable $\hat{O}$ diagonal in this basis
\begin{equation}
	\hat{O}=\sum_{k=1\ldots N}{O_k\roa{k}}.
\end{equation}
Moments of observable are calculated by probability distribution 
\[\aver{\hat{O}^n}=\tr{\hat{O}^n\hat{\rho}}=\sum_{k=1\ldots N}{O_k^n\mess{k}{\hat{\rho}}}=\sum_{k=1\ldots N}{O_k^np_k}.\]
In the case of values of $N-1$ independent moments being known (those are independent if all the eigenvalues of matrix of observable are different) one can calculate the whole probability distribution. Number of such measurements, more exactly dependence of accuracy of evaluation of probabilities or moments of observable is determined by standard methods of mathematical statistics.

Thus each observable with nondegenerate eigenvalue system creates basis of state space associated with that observable and set of projectors to eigenstates characterizing properties of states reduced in the process of given observable measurement.

\subsection{Ladder operators}
Finite-dimensional state spaces with practically arbitrary dimensions in quantum-mechanical problems are well known -- those are irreducible  representations of isotropy group of three-dimensional space characterized with given value of angular momentum. Eigenvectors of operator of angular momentum to given axis are created as result of sequential effect of one of special operators $\hat{L}_{\pm}$ on annihilation vector of second one of the pair of operators. 

Useful for future study property of those operators is the fact that such connection between the basis of state space and operator pair can be built for arbitrary basis. 

Let us denote $\left\{\cat{k}: k=1\ldots N\right\}$ an ordered in arbitrary way basis of $N$-dimensional state space of quantum system under consideration. For each basis one can define operator pair $\hat{L}_{\pm}$, ordering sequence of basis vectors
\begin{equation}\label{ladder_def}
\begin{array}{l}
	\hat{L}_+\stackrel{Def}{=} \sum^{N}_{k=1}{\sqrt{\left(N-k\right)k}\cat{k+1}\otimes\bra{k}}\\
 \hat{L}_-\stackrel{Def}{=} \sum^{N}_{k=1}{\sqrt{\left(N+1-k\right)\left(k-1\right)}\cat{k-1}\otimes\bra{k}}
\end{array}.
\end{equation}
Result of operator $\hat{L}_{+}$ effect on arbitrary vector $\cat{k}$ is next vector $\cat{k+1}$ or zero (if the vector has had the largest number). Similarly, result of effect of operator $\hat{L}_{-}$ on arbitrary vector  $\cat{k}$ is previous vector $\cat{k-1}$ or zero (if the vector has had the smallest number). Operator 
\begin{equation}
 \hat{L}_3\stackrel{Def}{=} \big(\hat{L}_+\hat{L}_--\hat{L}_-\hat{L}_+\big)/2= \sum^{N}_{k=1}{\left(k-1-N/2\right)\cat{k}\otimes\bra{k}},
\end{equation}
has basis vectors as eigenvectors. Re-denotion $\cat{k}\rightarrow\cat{m=k-1-N/2}: m=-N/2\ldots N/2$ adds to analogy between arbitrary basis and basis of irreducible representation of group of anisotropy. 

It is important to point out that operators $\hat{L}_{\pm}$ are built for each orthogonal basis irrespectively from the fact if that basis is related to isotropy group of real space or not. For each basis those operators give sequence of basis vectors, i.e. method of basis ranging. If from physical considerations it turns out to be needed to change the sequence of basis vectors one has to change basis ladder operators simultaneously -- ladder operators are associated to the basis of state space.

Ranging operators form basis of matrix algebra in state space. Since observables are given by Hermite matrices, arbitrary observable has representation through normally ranged combination of ladder operators
\begin{equation}
	\hat{O}=\sum_{m,n=1}^N{O_{m,n}\hat{L}_+^m\hat{L}_-^n};\ O_{m,n}=O_{n,m}^*.
\end{equation}
Each observable that is diagonal in chosen basis has eigenvalues $O_k$, interpolated by some function $o\left(x\right)$ of eigenvalues $k-1-N/2$ of operator $\hat{L}_3$ and thus is itself same function of that operator
\begin{equation}
	\hat{O}=o\left(\hat{L}_3\right);\ \mapsto 
	\hat{O}=\sum_{\forall k}{o\left(k-1-N/2\right)\cat{k}\otimes\bra{k}}.
\end{equation}

\subsection{Analyzers}
Result of interaction of the system being measured with analyzer is reduction of density matrix of the system to mix of states determined by analyzer. 

Formally analyzer is given by system of vectors on which states are projected after interaction with analyzer, or system of ladder operators related to those vectors. It is important that the system of analyzer vectors can be ranged in arbitrary way, and  own set of ladder operators corresponds to each ranging.

After having chosen analyzer, result of measurement is in distribution of probabilities $\left\{p_k: k=1\ldots N\right\}$ giving $N-1$ real numbers -- independent characteristics of density matrix of the state being measured, or $N-1$ moments of observable $\hat{L}_3$. For that observable, like any other observable, $N$-th moment depends on the previous ones, thus such result corresponds to complete measurement for the observable.

Since arbitrary density matrix has $N^2-1$ independent real parameters, one series of measurements can not define the state of the system being studied, one needs $N+1$ measurement series that use different in principle bases, i.e. ones that do not turn to each other by simple vector permutation. For such bases operators  $\hat{L}_3^{\left\{m\right\}}$ have not to commutate one with another. It is easy to build the sequence of mutually non-commutative operators having set the sequence of angles $\phi^{\left\{m\right\}}=m\pi/(N+1)$. Sequence of $N+1$ operators 
\begin{equation}
	\hat{L}_3^{\left\{m\right\}}
=\cos{\phi^{\left\{m\right\}}}\hat{L}_3
+\sin{\phi^{\left\{m\right\}}}\left(\hat{L}_++\hat{L}_-\right)/2:\ m=0\ldots N,
\end{equation}
noncommuting each one with each other
\begin{equation}
	\Big[\hat{L}_3^{\left\{m\right\}}\hat{L}_3^{\left\{n\right\}}\Big]=
	\sin{\phi^{\left\{m-n\right\}}}\left(\hat{L}_+-\hat{L}_-\right)/2,
\end{equation}
has as eigenvectors sequence of needed bases, and results of measurements for each of observables $\hat{L}_3^{\left\{m\right\}}$ determine additional $N-1$ real numbers -- independent characteristics of density matrix of the state being measured. Total number of values being measured coincides with the number of real parameters of density matrix, and in such way complete information on statistical properties of quantum information transfer channel is obtained.

The described scheme gives realization of the method of quantum tomography in application to arbitrary quantum system with finite-measure state space. 

Considerations given above make evidence of fact that complete measurement of properties of quantum state is impossible without measurement of incompatible observables. 
 This results in impossibility of realization of device able to clone state of quantum channel, thus in impossibility of exact copying of states of arbitrary quantum information transfer channel. An exception is a channel for which it is known beforehand what particular set of basis states is used in the process of state preparing. Just such channels with preliminary selection of states correspond to notion of properties of classical information transfer channels.

\section{Two-particle quantum channels} 
Two-component system usually consists of pair of qubits, in general case each of particles can have basis of state space with more than two vectors.  

The most often used examples are paraqubit and paraqutrit -- combination of qubit and qutrit.

\subsection{Description of state space of a pair}
We denote as $\mathcal{H}_A$ state space of particle of sort $A$ and $\mathcal{H}_B$ state space of particle of sort $B$, then common state space is direct product of those subspaces, $ \mathcal{H}=\mathcal{H}_A\otimes\mathcal{H}_B $.

For qubit pair common space is four-dimensional, $C^4=C^2\otimes C^2$, while for combination of qubit and qutrit -- six-dimensional, $C^6= C^2\otimes C^3$.
\subsubsection{Induced and entangled bases}
Induced basis of common state space$\left\{\catt{k}\in  \mathcal{H}\right\}$ is given by direct basis product $\left\{\cat{m}\in  \mathcal{H}_A\right\}$ of state space of particle of sort $A$  and $\left\{\catr{n}\in  \mathcal{H}_B\right\}$ of sort $B$ \begin{equation}
	\catt{k=m\cdot n}=\cat{m}\otimes\catr{n}\  \forall m,n.
\end{equation}
Induced basis of paraqubit state space is formed by four products
\begin{equation}
	\catt{0,0}=\cat{0}\otimes\catr{0},\ 	\catt{1,0}=\cat{1}\otimes\catr{0},\ 
	\catt{0,1}=\cat{0}\otimes\catr{1},\ 	\catt{1,1}=\cat{1}\otimes\catr{1}.
\end{equation}
Induced space of three qubit state space is formed by six products
\begin{equation}\begin{array}{lll}
	\catt{0,0}=\cat{0}\otimes\catr{0},& 
	\catt{0,1}=\cat{0}\otimes\catr{1},&
		\catt{0,2}=\cat{0}\otimes\catr{2},\\ 
	\catt{1,0}=\cat{1}\otimes\catr{0},& 
	\catt{1,1}=\cat{1}\otimes\catr{1},&
		\catt{1,2}=\cat{1}\otimes\catr{2}.
	\end{array}
\end{equation}

Common basis is result of arbitrary unitary transformation of induced basis. At least part of vectors of common basis is some linear combination of direct products and corresponds to entangled states.
\begin{equation}\label{arbasis}
	\catt{k}=\sum_{\forall m, \forall n}{c^{\left\{k\right\}}_{m,n}\cat{m}\otimes\catr{n}}.
\end{equation}
Such entangled basis is to be used, for instance, in the case of its vectors being eigenvectors of density matrix.

If the particles are identical it is needed to use (anti)symmetrical basis vectors. Under such condition in the basis of qubit state space either the triplet system of symmetrical states
\begin{equation}\begin{array}{l}
	\catt{0,0}=\cat{0}\otimes\catr{0},\\ 	\catt{1,1}=\cat{1}\otimes\catr{1},\\ 
	\catt{0,1}=\frac{1}{\sqrt{2}}\Big(\cat{0}\otimes\catr{1}+\cat{1}\otimes\catr{0}\Big),
	\end{array}
\end{equation}
or singlet antisymmetrical state
\begin{equation}
	\catt{0,1}=\frac{1}{\sqrt{2}}\Big(\cat{0}\otimes\catr{1}-\cat{1}\otimes\catr{0}\Big).
\end{equation} 
is left.
The last state space is one-dimensional.
\subsubsection{Matrices of observables}
Observables of composite system are given by Hermitian matrices in composite basis
\begin{equation}
	\hat{O}=\sum_{\forall k,kk}{O_{k,kk}\catt{k}\brat{kk}},
\end{equation}
and can be given by Hermitian matrices in induced basis as well
\begin{equation}
	\hat{O}=\sum_{\forall m,mm;n,nn}{O_{m,mm;n,nn}\cat{m}\bra{mm}\otimes\catr{n}\brar{nn}}.
\end{equation}
In this basis matrices of observables for each subsystem have simplified form. 

\subsubsection{Eigenbasis of observable}
Hermitian matrix of each observable has system of eigenvectors and eigenvalues.  
\begin{equation}
	\hat{O}\catt{k}=O_k\catt{k}.
\end{equation}
In the case of eigenvalues being non-degenerate, system of eigenvectors is defined uniquely and forms orthogonal representation of unity 
\begin{equation}
	\sum_{\forall k}{\catt{k}\brat{k}}=\hat{1}.
\end{equation}
For degenerate system of eigenvalues system of eigenvectors is defined ambiguously, though possibility to choose an orthogonal basis remains.
\subsubsection{Observables of subsystems}
Observables depending only on subsystem $A$ or $B$, respectively,  have as their matrices expressions
\begin{equation}\label{O_subs}
\begin{array}{lccc}
	\hat{O}_A
=&\sum_{\forall m,mm}{O_{m,mm}\cat{m}\bra{mm}}&\otimes&\hat{1}_B;\\
	\hat{O}_B
=&\hat{1}_A&\otimes &\sum_{\forall n,nn}{O_{n,nn}\catr{n}\brar{nn}}.
\end{array}
\end{equation}
For pure states belonging to induced basis average values of observables of one subsystem do not depend on properties of state of the other subsystem
\begin{equation}
\begin{array}{ll}
	\mest{m,n}{\hat{O}_A}
=&O_{m,m};\\
	\mest{m,n}{\hat{O}_B}
=&O_{n,n}.
\end{array}
\end{equation}
Eigenvectors of observables of a subsystem are always degenerate since, for instance, in the case of some vector $\cat{\psi}\otimes\catr{\xi}$ being eigenvector for observable of subsystem $A$, i.e. $\hat{O}_A\cat{\psi}\otimes\catr{\xi}=O\cat{\psi}\otimes\catr{\xi}$, vector $\cat{\psi}\otimes\catr{\zeta}$ is eigenvector with the same eigenvalue $\hat{O}_A\cat{\psi}\otimes\catr{\zeta}=O\cat{\psi}\otimes\catr{\zeta}$ as well. 

Projection of observables of subsystem to subsystem state spaces is performed in quite simple way, it is enough to remove unit matrix of other subsystem in the expression \eref{O_subs} 
\begin{equation}\label{O_prosubs}
\begin{array}{lc}
	\hat{O}_A|_A
=&\sum_{\forall m,mm}{O_{m,mm}\cat{m}\bra{mm}};\\
	\hat{O}_B|_B
= &\sum_{\forall n,nn}{O_{n,nn}\catr{n}\brar{nn}}.
\end{array}
\end{equation}
Eigenvectors of such matrices create subsystem state space bases associated with respective observables, and with those -- induced basis of composite system state space. 
 
Eigenvectors of arbitrary observable not always coincide with vectors of induced basis.

Paraqubit has as observables of subsystems
\begin{equation}
\begin{array}{lccc}
	\hat{O}_A
=&\sum_{m,mm=0,1}{O_{m,mm}\cat{m}\bra{mm}}&\otimes&\hat{1}_B.\\
	\hat{O}_B
=&\hat{1}_A&\otimes &\sum_{n,nn=0,1}{O_{n,nn}\catr{n}\brar{nn}}.
\end{array}
\end{equation}

\subsubsection{Density matrix}

Density matrix of composite system like matrices of observables is Hermitian one, though it has two specific peculiarities: first, density matrix has nonnegative eigenvalues only, $\left\{0 \leq \rho_k\leq 1:\ \forall k  \right\}$, and second, the sum of eigenvalues of density matrix is equal to unity, $\sum_{\forall k}{\rho_k}=1$.

From these properties follows possibility to treat the values of density matrix as probabilities of quantum system states corresponding to eigenvectors and to give to density matrix expansion by eigenvalues 
\begin{equation}
	\hat{\rho}=
		\sum_{\forall k}{\rho_{k}\catt{k}\brat{k}},
\end{equation}
treatment of probability of pure state mix $\catt{k}$, in which relative part of each state is defined by respective eigenvalue of density matrix $\rho_k$. 

Eigen basis of density matrix is usually formed by vectors of state space of composite system that are nontrivial combinations  \eref{arbasis} of vectors of induced basis, and representation of density matrix in induced basis has not only diagonal components
\begin{equation}\label{red_rho}
	\hat{\rho}=
		\sum_{\forall m,m'; \forall n,n'}{\Big[\sum_{\forall k}{\rho_{k}c^{\left\{k\right\}}_{m,n} c^{*\left\{k\right\}}_{mm,nn}}\Big]
	\left(\cat{m}\otimes\bra{mm}\right)\otimes
	\left(\catr{n}\otimes\brar{nn}\right)}.
\end{equation}
Density matrices of subsystems are built through averaging density matrix of composite system by states of the second subsystem
\begin{equation}\label{sub_rhos}
	\hat{\rho}^{\left\{A\right\}}=\sum_{\forall n}{\brar{n}\hat{\rho}\catr{n}};\ 
	\hat{\rho}^{\left\{B\right\}}=\sum_{\forall m}{\bra{m}\hat{\rho}\cat{m}}.
\end{equation}
With account of explicit form of density matrix in induced basis \eref{red_rho} we have
\begin{equation}
\begin{array}{l}
	\hat{\rho}^{\left\{A\right\}}=
	\sum_{\forall n, k ;m,mm}{\rho_{k}c^{\left\{k\right\}}_{m,n} c^{*\left\{k\right\}}_{mm,n}\cat{m}\bra{mm}};\\ 
	\hat{\rho}^{\left\{B\right\}}=
\sum_{\forall m, k ;n,nn}{\rho_{k}c^{\left\{k\right\}}_{m,n} c^{*\left\{k\right\}}_{m,nn}\catr{n}\brar{nn}}	.
\end{array}
\end{equation}
Usually in superposition \eref{arbasis} to one basis vector of one particle only one basis vector of other one corresponds, in such case density matrices of subsystems are diagonal
\begin{equation}
\begin{array}{l}
	\hat{\rho}^{\left\{A\right\}}=
	\sum_{\forall m}{\Big[\sum_{\forall n, k}{\rho_{k}\abs{c^{\left\{k\right\}}_{m,n}}^2}\Big] \cat{m}\bra{m}}=
	\sum_{\forall m}{\rho^{\left\{A\right\}}_m \cat{m}\bra{m}};\\ 
	\hat{\rho}^{\left\{B\right\}}=
\sum_{\forall n}{\Big[\sum_{\forall m, k}{\rho_{k}\abs{c^{\left\{k\right\}}_{m,n}}^2}\Big]\catr{n}\brar{n}}=
	\sum_{\forall n}{\rho^{\left\{B\right\}}_n \catr{n}\brar{n}}.	
\end{array}
\end{equation}

Depending on properties of eigen basis density matrix of composite system in specific cases can be product of density matrices of subsystems, or mix of such products -- both those cases can take place for classical paired information transfer channel as well. Quantum properties of paired channel, especially ones that have no classical analog, take place in all the cases of channel state being irreducible to some classical variant, like entangled states of paraqubit being irreducible to classical analog. Non-classic states, similarly to the case of paraqubit, are called entangled states, that's why all the states of two-particle quantum information transfer channel  can be divided into following three types:
\begin{enumerate}
	\item Independent subsystems: $\hat{\rho}=\hat{\rho}_A\otimes\hat{\rho}_B$;
	\item Mix of independent subsystems: $\hat{\rho}=  \sum_{s}{p_s\hat{\rho}^{\left(s\right)}_A \otimes\hat{\rho}^{\left(s\right)}_B}$;
	\item Entangled: all the others.
\end{enumerate}

\subsection{Measurement of states of two-particle channel}
The most important distinction of the process of measurement of two-particle quantum information transfer channel states is in possibility of simultaneous measurement of two observables. In the case of those observables being one-particle one can distinguish by results of measurements states in which separate particles form independent subsystems -- in such states results of measurement of single-particle observables are independent. In the states that are mixes of subsystems, as well as in entangled states, results of measurement of one-particle observables correlate. 

Real characteristics of the state of particles and the channel in whole are not values of observables but probability distributions in given set of detectors. Those probability distributions can be determined directly by statistics of sequential measurements or by means of needed number of moments of some observable.
 Each set of one-particle detectors has as its mathematical model sequence of projectors to eigenstates of one-particle observable. If eigenvalues of observable are non-degenerate its own states form basis, while detectors, like  in \eref{det_basis}, are simulated by projectors to basis states 
\begin{equation}
	\hat{P}^{\left\{A\right\}}_{m}=\roa{m};\ \hat{P}^{\left\{B\right\}}_{n}=\eroa{n}.
\end{equation}
Complete system of detectors of composite system is simulated by direct products of such projectors
\begin{equation}
	\hat{P}_{m,n}=\hat{P}^{\left\{A\right\}}_{m}\hat{P}^{\left\{B\right\}}_{n}
	=\roa{m}\otimes\eroa{n}=\proa{m,n}.
\end{equation}
To simultaneous measurement of both particles simultaneous operation of some $m$-th detector $A$ and $n$-th detector $B$ corresponds. Result of such measurements is distribution of probabilities of simultaneous operation of detectors of both types, this can be calculated by density matrix of composite system as trace of product of density matrix and both projectors
\begin{equation}
	P_{m,n}=\tr{\hat{\rho}\hat{P}^{\left\{A\right\}}_m\hat{P}^{\left\{B\right\}}_n}.
\end{equation}
\subsubsection{Measurement of independent states}
In the case of density matrix of system being product of density matrices of subsystems the trace is divided into product of two traces
\begin{equation}
	\tr{\hat{\rho}_{A}\otimes\hat{\rho}_{B} \hat{P}^{\left\{A\right\}}_m\hat{P}^{\left\{B\right\}}_n}
=\tr{\hat{\rho}_{A}\hat{P}^{\left\{A\right\}}_m}
\cdot \tr{\hat{\rho}_{B}\hat{P}^{\left\{B\right\}}_n}.
\end{equation}
Probability distributions for particles are respectively independent 
\begin{equation}
	P_{m,n}=P_{m}^{\left\{A\right\}}\cdot P_{n}^{\left\{A\right\}}.
\end{equation}
So, two-particle quantum information transfer channel with independent particles creates independent subchannel state probability distributions, like classical parallel channel. 

Two other types of two-particle channel states create correlated probability distributions.

\subsubsection{Classical correlations}
For states that are classical mix of independent states correlation of results of measurements is specific. Classical correlation of particle -- subsystem states is worthy of notice because more important, exceptionally classical correlations are masked by it. We restrict our consideration by the simplest example of such states -- mix of two qubits with diagonal density matrix. Let us suppose that with probability $p$ both qubits are in zero state, and with probability $1-p$ -- in state one
\begin{equation}
	\hat{\rho}=p\cat{0}\bra{0}\otimes\catr{0}\brar{0} +\left(1-p\right)\cat{1}\bra{1}\otimes\catr{1}\brar{1}.
\end{equation}
Density matrix of each particle is calculated by averaging by states of other particle and looks like
\begin{equation}
	\hat{\rho}^{\left\{A\right\}}=p\cat{0}\bra{0} +\left(1-p\right)\cat{1}\bra{1};\ 
	\hat{\rho}^{\left\{B\right\}}=p\catr{0}\brar{0} +\left(1-p\right)\catr{1}\brar{1}.
\end{equation}
Results of measurements for each particle are state 0 with probability $p$ and 1 with probability $1-p$. Along with that probability to get state 0 for one particle under condition that the other one is registered in state 1, is equal to zero, thus states of particles are totally correlated. 

Statistic character of results of measurements for channel in the example given is formed on the step of state preparation since in series of measurements with length $N$ channel is prepared in state $\catt{0,0}$ $Np$ times, and in state$\catt{1,1}$ $N(1-p)$ times, respectively.

Entropy of state 
\begin{equation}
	S=-\tr{\hat{\rho}\log_2\hat{\rho}},
\end{equation}
in this example coincides with common entropy of two-level classical doubled channel. That is the value of entropy of each of the channels as well. 

Information obtained at each measurement of channel state, or states of particles, is equal to  $-\log_2p$ or $-\log_2(1-p)$,  respectively, so in such channels information put to channel is simply doubled for two particles, and process of measurement does not produce new information. 
 
\subsubsection{Production of information}

Correlation of probability distributions of measured states of separate particles of two-particle quantum channel is most important for cases when state of channel is pure one. Let us suppose that vector of channel state is some superposition of basis vectors induced by eigenstates of observables being measured
\[
	\catt{\psi}=\sum_{k}{\psi_k\cat{m_k}\otimes\catr{n_k}}:\ \sum_{k}{\abs{\psi_k}^2}=1.
\]
One can arrange states of particles in such way that numbers of states $m_k$, $n_k$ coincide with the number $k$ of term in superposition. 
\begin{equation}\label{pure_sup}
	\catt{\psi}=\sum_{k}{\psi_k\cat{k}\otimes\catr{k}}.
\end{equation}
State of each particle of channel is not pure. It is obtained from density matrix of pure state of composite system by averaging by states of the other particle
\begin{equation}
	\hat{\rho}^{\left\{A\right\}}=
	\sum_{\forall n}{\brar{n}\catt{\psi}\brat{\psi}\catr{n}}.
\end{equation}
After substitutions and simplifications 
\[		\hat{\rho}^{\left\{A\right\}}=\sum_{\forall n}{\brar{n}\sum_{k,kk}{
	\psi_k\psi^*_{kk}\cat{k}\bra{kk}\catr{k}\brar{kk}
	}\catr{n}}
\]
\[=\sum_{\forall n;k,kk}{\brar{n}\catr{k}
	\psi_k\psi^*_{kk}\cat{k}\bra{kk}\brar{kk}\catr{n}}
=	\sum_{\forall n;k,kk}{\delta_{n,k}
	\psi_k\psi^*_{kk}\cat{k}\bra{kk}\delta_{kk,n}}
\]
we get density matrices of particle states
\begin{equation}
\hat{\rho}^{\left\{A\right\}}=
	\sum_{k}{\abs{\psi_k}^2\cat{k}\bra{k}};\ 
		\hat{\rho}^{\left\{B\right\}}=
	\sum_{k}{\abs{\psi_k}^2\catr{k}\brar{k}}.
\end{equation}
Since those states are mixed ones they have entropy of entanglement
\begin{equation}
	S^{\left\{A,B\right\}}=-\sum_{k}{\abs{\psi_k}^2 \log_2 \abs{\psi_k}^2},
\end{equation}
that can become zero only in the case if in superposition \eref{pure_sup} only one coefficient differs from zero (it is then equivalent to one).

Let us now consider information evaluation of the process of measurement. Since the states of particles are registered with probabilities $p_k=\abs{\psi_k}^2$, information obtained at next measurement for particle of sort $A$ or $B$ in state $k$, is equal to $I_k=-\log_2 p_{k}=-\log_2 \abs{\psi_k}^2$. Those values are in full agreement with the value of entropy of channels.

Let us now consider joint probability of registration of $m$-th state of particle of $A$-th sort and $n$-th state of particle of $B$-th sort. It is calculated as average value of product of two projectors 
\begin{equation}
	p_{m,n}=\tr{\catt{\psi}\brat{\psi}\cat{m}\bra{m}\catr{n}\brar{n}},
\end{equation}
and is equal 
\begin{equation}
	p_{m,n}=\sum_{k}{\abs{\psi_k}^2\delta_{n,k}\delta_{k,m}}
	=p_m \delta_{n,m}.
\end{equation}
For each value $m$ in this sum only one term with given number $k$ is left, and $n$ is to correspond to the same number $k$. So, measurements of states of particles are totally correlated -- measurement of $k$-th state of one particle is necessarily followed by measurement of the same $k$-th state of another one.

Pure states differ by fact that in the process of formation of those no information is placed to channel -- the channel is prepared in one pure state for each next attempt of measurement. As consequence, statistic deflections of results of measurements from trial to trial give information that has not been entered to channel, this information is produced in the process of measurement of the channel. Cause of information production in each measurement is in non-compatibility (noncommutativity of respective matrices) of analyzer preparing the state for transfer in channel with the analyzer performing state reduction before detection. Just in the process of channel state reduction by analyzer of measuring device new information is produced.

\subsection{Degeneration of states}
If correlations between particles of two-particle quantum information transfer channel are to be classical or quantum ones, depends on properties of eigenvalues and eigenvectors of density matrix, though that dependence is not very much transparent. 

Channel with independent particles density matrix of which has as eigenvectors products of eigenvectors of subchannels, and eigenvalues -- products of eigenvalues is easily recognized.
\begin{equation}
	\hat{\rho}^{\left\{A\right\}}\otimes\hat{\rho}^{\left\{B\right\}}
	\cat{m}\otimes\catr{n}=
	\rho^{\left\{A\right\}}_m\rho^{\left\{B\right\}}_n\cat{m}\otimes\catr{n}.
\end{equation}

In same easy way properties of pure state are determined -- such state either has independent particles, or it is entangled. 

The most complicated is the boundary between entangled states and mixes of independent states since mix of entangled states can have properties same to that of the mix of independent ones. Example of such mix of entangled states is equilibrium mix of states $\catt{\pm}=\frac{1}{\sqrt{2}}\big(\catt{0,0}\pm\catt{11}\big)$. 
Density matrices of each state
\[\hat{\rho}_{\pm}=\frac{1}{2}\Big(\proa{0,0}+\proa{1,1}\Big)\pm
\frac{1}{2}\Big(\catt{11}\brat{00}+\catt{00}\brat{11}\Big),
\] 
include diagonal components and interference terms, while equilibrium mix of those 
\[\hat{\rho}_{1/2}=\frac{1}{2}\Big(\hat{\rho}_{+}+\hat{\rho}_{-}\Big)
=\frac{1}{2}\Big(\proa{0,0}+\proa{1,1}\Big),
\] 
has diagonal terms only, and again is equilibrium mix of independent states \[\proa{0,0}=\roa{0}\otimes\eroa{0};\  \proa{1,1}=\roa{1}\otimes\eroa{1}.\]

This example makes evidence that only pure states can be divided into independent and entangled ones, while mix of entangled states can be completely equivalent to the mix of independent ones. Since entangled pure states differ from independent pure states by presence of quantum correlations, the question on effective criteria of existence or absence of quantum correlations in given mixed state remains open.

One should notice that transformation of mix of entangled states to mix of independent ones, at least in this example, is followed by degeneration of density matrix. Non-equilibrium mix of entangled states
\[\hat{\rho}_{p\neq1/2}=p\hat{\rho}_{+}+\left(1-p\right)\hat{\rho}_{-}=\frac{1}{2}\hat{1}+
(p-1/2)\Big(\catt{11}\brat{00}+\catt{00}\brat{11}\Big),
\] 
has density matrix in which interference terms remain non-compensated, those provide correlation between particles of two-particle quantum information transfer channel. As to such mix, problem on possibility or impossibility of existence of such correlations in mix of independent particles remains unresolved.

\section{Classification of quantum particle pair}
Division of states of particle pair by correlation properties of measurement results is technically complicated -- to find out to which type does specific state belong one has to consider an enough (and what is enough?) number of needed pairs of observables (and which ones are needed?) and to carry out for each pair a needed quantity of measurements, and after that one can consider to which type does the state belong. More effective is to combine correlation properties of states of particle pair to other ones, like state symmetry relatively to group of symmetry representing properties of quantum channel of information transfer.  
\subsection{Evolution of states}
It is usually supposed that quantum channel of information transfer is a closed quantum system, i.e. environment does not affect, or does not make considerable effect on processes that take place in the channel. Under this supposition channel state at any instant remains pure state, if at having left the source it has been pure one, and this means that there exists a matrix of state evolution $\hat{U}\left(t\right)$ - matrix of transformation of initial channel state $\cat{in}$ to current one $\cat{t}$ which preserves norm and is unitary because of that
\begin{equation}
	\cat{t}=\hat{U}\left(t\right)\cat{in};\ \hat{U}^{+}\left(t\right)=\hat{U}^{-1}\left(t\right).
\end{equation}
Through adding to state transformation from initial instant with transformations from one instant to another 
\begin{equation}
	\cat{t}=\hat{U}\left(t,t_0\right)\cat{t_0};\ \hat{U}\left(t\right)=\hat{U}\left(t,t_0\right)\hat{U}\left(t_0\right),
\end{equation}
we get a group of matrices $\left\{\hat{U}\left(t,t_0\right): \forall t\geq t_0\right\}$, determining unitary transformations of channel states with time. This group has as generator Hermitian operator -- Hamiltonian of the channel 
\begin{equation}
	\hat{H}\left(t\right)=i\hbar\hat{U}^+\left(t\right)\frac{d}{dt}\hat{U}\left(t\right),
\end{equation}
creating equation of motion of particles of the channel -- Shr\"odinger equation
\begin{equation}
i\hbar\frac{d}{dt}\cat{t}=	\hat{H}\left(t\right)\cat{t}.
\end{equation}
Properties of Hamiltonian -- how in particular Hamiltonian acts on states of the channel, how it depends on physical properties of channel, are important at study of specific channels; in general case the fact of channel states transformations existence is enough. Under effect of that group channel state space is split to orbits - subspaces comprising vectors that turn to each other under effect of the group. One should notice that for independent from time Hamiltonian all orbits of the group are one-dimensional. In each orbit own irreducible representation of evolution group acts, channel properties same for all the states of one orbit do not depend on time and can be used for classification of channel states. 
\subsection{Internal symmetries}
The laws determining properties of quantum information transfer channel do not depend on the fact of what particular basis of channel state space is used for description of vectors, matrices or observables. That's why those laws are to be written in form invariant relatively to unitary transformations of state space basis. This statement is to be strengthened at consideration of two-particle quantum information transfer channel since state space of such channel $\mathcal{H}=\mathcal{H}_A\otimes\mathcal{H}_B$ is direct product of state space of one-particle subchannels, and all the laws are to be invariant with respect to unitary transformations of state space basis for each channel separately. Hereinafter dimensions of subchannel state spaces are denoted $N_{A,B}=\dim \mathcal{H}_{A,B}$, dimension of channel state space is equal to $N=N_{A}\cdot N_{B}$, respectively. 

Because of existence of groups of transformation of subchannel state space bases  $U\left(N_{A,B}\right)$ channel states can be classified by irreducible representations of any of those groups. Such classification has to rest on specific representation of the group of symmetry of subchannel state space in the group of state spaces of composite channel $U\left(N_{A}\right)\rightarrow U\left(N\right)$ and the second channel $U\left(N_{A}\right)\rightarrow U\left(N_{B}\right)$. The first representation can be exact since it is embedded, and the second one is exact only in the case of $N_{A}\leq N_{B}$. Here we consider that the subchannel $B$ can have higher dimensionality of state space. 

Practically complete classification of state space can be achieved with use of even not complete group of unitary transformations for the smaller of subchannel state spaces, but of its subgroup $U\left(2\right)$ only; this subgroup is complete for qubit state space only, though it always remains a subgroup of group of unitary transformations of states with arbitrary dimensionality.

\subsection{Basis transformation group}
Let us consider the basis of state space of composite channel \eref{para_pure} induced by subchannel state space bases and suppose that  specific representation of group $U\left(2\right)$ by transformations of states of each subchannel is given; then induced basis is transformed by direct product of those representations. 
\begin{equation}\label{sym_def}
	g\in U\left(2\right) \ \Rightarrow \ \cat{m} \rightarrow \hat{g}_A\cat{m};\ \catr{n} \rightarrow \hat{g}_B\catr{n}\ \Rightarrow \ \catt{k} \rightarrow \hat{g}\left(\equiv\hat{g}_A\otimes\hat{g}_B\right)\catt{k}.
\end{equation}
It is important to point out that symmetry of subchannel states with respect to $U\left(2\right)$ results from the fact that subchannel state spaces have dimensionality not smaller than two only. This takes place not only for spin subsystem of electron states, or polarization -- of photon ones, but for states of two-level atom as well.
\subsubsection{Transformation of ladder operators of one system}
With each basis of state space set of ladder operators  \eref{ladder_def} is associated. Unitary transformation of basis creates linear transformation of ladder operators belonging to one group of orthogonal transformations of three-dimensional vector space
\begin{equation}\label{ladder_rep}
g\in U\left(2\right) \ \Rightarrow \ 
   \hat{g}\hat{L}_a\hat{g}^+ =\sum_{b=1,2,3} C_a^b\left(g\right)\hat{L}_b,
\end{equation}
that belongs to the group of orthogonal transformations of three-dimensional vector space $SO(3)$.  With respect to this representation of internal group of symmetry ladder operators are vectors that form basis of vector space. Arbitrary orthonormalized basis of that space is a set of ladder operators to which specific basis of state spaces corresponds.
\subsubsection{Transformation of ladder operators of composite system}
Extension of group of basis transformations to state space of composite system \eref{sym_def} produces similar to \eref{ladder_rep} transformations of ladder operators of the system as a whole, along with both subsystems. 

Let us denote as $\hat{J}_a$, $\hat{S}_a$ and $\hat{L}_a$ ladder operators associated to basis of composite channel state space $\left\{\catt{k}:\ k=1\ldots N\right\}$, basis of state space of the smaller subchannel $\left\{\cat{m}:\ m=1\ldots N_A\right\}$ and basis of state space of larger subchannel $\left\{\catr{n}:\ n=1\ldots N_B\right\}$, respectively.
Among all the representations of group $U\left(2\right)$ in subchannel state spaces there exists such a one that each transformation of basis under effect of element of group is accompanied by transformation of ladder operators that belongs to the group of orthogonal transformations $SO(3)$ of three-dimensional vector space
\begin{equation}\label{ladders_ort}
g\in U\left(2\right) \ \Rightarrow \ {\begin{array}{ll}
   \hat{g}_A\hat{S}_a\hat{g}_A^+ &=\sum_{b=1,2,3} C_a^b\left(g\right)\hat{S}_b\\
   \hat{g}_B\hat{L}_a\hat{g}_B^+ &=\sum_{b=1,2,3} C_a^b\left(g\right)\hat{L}_b\\
   \hat{g}_A\otimes \hat{g}_B\hat{J}_a\hat{g}_B^+\otimes \hat{g}_A^+  &=\sum_{b=1,2,3} C_a^b\left(g\right)\hat{J}_b\\
\end{array}	}
\end{equation}
One can substantiate choice of representation \eref{ladders_ort} of the group $U\left(2\right)$ of symmetry of state subspaces by the group of orthogonal transformations of vector space  $SO(3)$ in another way as well. In vicinity of unit of group $U\left(2\right)$ result of transformation of ladder operators is a linear combination of ladder operators in form \eref{ladders_ort}, since Lie algebra of group $U\left(2\right)$ has representation by Lie algebra $SO(3)$. Expressions \eref{ladders_ort} extend this algebra representation to group representation.

\subsubsection{Irreducible representations}
Analogy of algebra of ladder operators to Lie algebra $SO(3)$ is extended to analogy to rules of angular momentum adding since ladder operators of subsystems are similar to spin $\hat{S}_a$, and orbital $\hat{L}_a$ parts of momentum $\hat{J}_a=\hat{L}_a+\hat{S}_a$. This last analogy results in possibility of classification of states of composite system by means of irreducible representations of the group.

Each irreducible representation of composite system has basis comprising eigenvectors of operators  $\hat{J}_3$ and $\hat{J}^2= \hat{J}^2_3+\hat{J}_3+\hat{J}_+\hat{J}_-$. Each of those basis vectors is some superposition of eigenstates of ladder operators of each of subsystems. 

Now we use denotations  $l=\left(N_B-1\right)/2$ and $s=\left(N_A-1\right)/2 \le l$. Each of irreducible representations of composite system is eigen subspace of operator  $\hat{J}^2= \hat{J}^2_3+\hat{J}_3+\hat{J}_+\hat{J}_-$ with eigenvalue from sequence  $j=l-s,l-s+1,\ldots l+s$. Each of those has dimensionality $N_j=2j+1$, and eigen vectors $\catt{j,m_j}$ of operator $\hat{J}_3$
\begin{equation}
	\hat{J}_3\catt{j,m_j}=m_j\catt{j,m_j},
\end{equation}
 have representations through products of eigenvectors $\cat{l,m_l}\otimes\catr{s,m_s}$ of operators $\hat{L}_3$ and $\hat{S}_3$
\begin{equation}\label{clebsh}
	\catt{j,m_j}=\sum_{m_s=-s\ldots s}{
	C_{j,m_j;m_s}\cat{l,m_l=m-m_s}\otimes\catr{s,m_s},
	}
\end{equation}
this is known as law of momentum addition. In this sum products of state vectors of particles with sum of eigenvalues $m_l$, $m_s$ of ladder operators$\hat{L}_3$ and $\hat{S}_3$ equal to eigenvalue $m_j$ of operator $\hat{J}_3$ take part.

Coefficients of the sum $C_{j,m;m_s}$ are determined by Clebsch-Gordan $3j$-symbols  
\begin{equation}
	C_{j,m;m_s}=\left(-1\right)^{l-s+m}\sqrt{2j+1}\Big(
	\begin{array}{ccc}
	l&s&j\\m_s-m&m_s&-m
	\end{array}
	\Big)
\end{equation}
Coefficients of the sum  $C_{j,m_j;m_s}$ differ from zero for all the values of projection of momentum of the smaller subsystem for which difference $m_l=m_j-m_s$, that gives projection of momentum of the second subsystem is within $-l\leq m_l \leq l$. As the result, only two boundary states 
\begin{equation}\label{dist_state}
	\catt{l+s,\pm\left(l+s\right)}=\cat{l,\pm l}\otimes\catr{s,\pm s},
\end{equation}
are direct products of one-particle states, while all the other basis vectors of irreducible representations \eref{clebsh} correspond to entangled states.

\paragraph{Basis of paraqubit irreducible representations}
Paraqubit has state spaces of particles with same dimensionality 2. Parameter $s=l$ is equal to $1/2$, two irreducible representations $j=1/2-1/2=0$ and $j=1/2+1/2=1$ exist. Dimensionality of the space of the first representation is 1, of the second one -- 3. 

State space basis of one-dimensional (singlet) representation forms state vector
\begin{equation}\label{para_s}
	\catt{s}=\frac{1}{\sqrt{2}}\cat{\frac{1}{2}}\otimes\ecat{-\frac{1}{2}}
	-\frac{1}{\sqrt{2}}\cat{-\frac{1}{2}}\otimes\ecat{\frac{1}{2}},
\end{equation}
that is the first of known entangled states.

Three-dimensional (triplet) representation has as its basis the states
\begin{equation}\label{para_t}
\begin{array}{ll}
	\catt{1,-1}&=\cat{-\frac{1}{2}}\otimes\ecat{-\frac{1}{2}};\\
	 \catt{0}&=\frac{1}{\sqrt{2}}\cat{\frac{1}{2}}\otimes\ecat{-\frac{1}{2}}
	+\frac{1}{\sqrt{2}}\cat{-\frac{1}{2}}\otimes\ecat{\frac{1}{2}};\\
	\catt{1,1}&=\cat{\frac{1}{2}}\otimes\ecat{\frac{1}{2}}.
\end{array}
\end{equation}
Among those states two are independent, and only one is entangled.

Arbitrary linear combination of vectors \eref{para_t}
\[a\catt{1,-1}+b \catt{0}+c\catt{1,1}=\]
\[\Big(a\cat{-\frac{1}{2}}+\sqrt{\frac{1}{2}}b\cat{\frac{1}{2}}\Big)\otimes\ecat{\frac{1}{2}}
+\Big(\sqrt{\frac{1}{2}}b\cat{-\frac{1}{2}}+c\cat{\sqrt{\frac{1}{2}}b}\Big)\otimes\ecat{-\frac{1}{2}}
\]
is product of two vectors
\[\Big(a\cat{-\frac{1}{2}}+\sqrt{\frac{1}{2}}b\cat{\frac{1}{2}}\Big)\otimes\Big(\ecat{-\frac{1}{2}}+k\ecat{\frac{1}{2}}\Big)
\]
in the case of satisfaction of equations
\[a=k^2c;\ b=\sqrt{2}kc;\ c=\frac{1}{1+\abs{k}^2}.
\]
Set of such states is topologically equivalent to sphere, and opposite in diameter points of the sphere correspond to orthogonal states forming the basis of entangled triplet states.

Completely disentangled basis can be build only with use of vectors of both subspaces of irreducible representations.

\paragraph{Paraqutrit irreducible representation basis}
State space of channel is divided into two subspaces -- two-dimensional $j=1/2$ and four-dimensional $j=3/2$ ones. 

Two-dimensional subspace has total momentum $j=l-s=1-1/2$ smaller than sum of momenta, that's why to each value of total momentum projection two possible combinations of momentum components, $\cat{+1}\otimes\ecat{-\frac{1}{2}}$ and $\cat{\ 0}\otimes\ecat{+\frac{1}{2}}$ correspond, those together form the state $\pcat{\frac{1}{2},+\frac{1}{2}}$, while $\cat{\ 0}\otimes\ecat{-\frac{1}{2}}$ and $\cat{-1}\otimes\ecat{+\frac{1}{2}}$ form the state $\pcat{\frac{1}{2},-\frac{1}{2}}$:
\begin{equation}
	\begin{array}{c}
	\pcat{\frac{1}{2},+\frac{1}{2}}
		=\frac{1}{\sqrt{3}}\cat{\ 0}\otimes\ecat{+\frac{1}{2}}
		-\sqrt{\frac{2}{3}}\cat{+1}\otimes\ecat{-\frac{1}{2}}\\
		\pcat{\frac{1}{2},-\frac{1}{2}}
		=-\sqrt{\frac{2}{3}}\cat{-1}\otimes\ecat{+\frac{1}{2}}
		+\frac{1}{\sqrt{3}}\cat{\ 0}\otimes\ecat{-\frac{1}{2}}\end{array}	.
\end{equation}
Both basis vectors are entangled, and even arbitrary combination of those vectors does not degenerate to direct product.

In four-dimensional subspace two one-dimensional subspaces of independent states are chosen
\begin{equation}\label{three_dis}
	\begin{array}{c}
	\pcat{\frac{3}{2},+\frac{3}{2}}	=\cat{+1}\otimes\ecat{+\frac{1}{2}}\\
		\pcat{\frac{3}{2},-\frac{3}{2}}	=\cat{-1}\otimes\ecat{-\frac{1}{2}}
	\end{array},
\end{equation}
and two-dimensional subspace of entangled states
\begin{equation}\label{three_ent}
	\begin{array}{c}
	\pcat{\frac{3}{2},+\frac{1}{2}}
		=\frac{1}{\sqrt{3}}\cat{+1}\otimes\ecat{-\frac{1}{2}}
		+\sqrt{\frac{2}{3}}\cat{\ 0}\otimes\ecat{+\frac{1}{2}}\\
	\pcat{\frac{3}{2},-\frac{1}{2}}
		=\frac{1}{\sqrt{3}}\cat{-1}\otimes\ecat{+\frac{1}{2}}
		+\sqrt{\frac{2}{3}}\cat{\ 0}\otimes\ecat{-\frac{1}{2}}\end{array}.
\end{equation}
In four-dimensional subspace remains not entangled \eref{three_dis} or linear combinations of all four basis vectors.

Arbitrary linear combination of vectors \eref{three_dis} and \eref{three_ent}
\[a\catt{3/2,3/2}+b\catt{3/2,1/2}+c\catt{3/2,-1/2}+d\catt{3/2,-3/2}=\]
\[\Big(a\cat{1}+\sqrt{\frac{2}{3}}b\cat{0}+\sqrt{\frac{1}{3}}c\cat{-1}\Big)\otimes\ecat{1/2}
+\Big(\sqrt{\frac{1}{3}}b\cat{1}+\sqrt{\frac{2}{3}}c\cat{0}+d\cat{-1}\Big)\otimes\ecat{-1/2}
\]
is product of two vectors
\[\Big(a\cat{1}+\sqrt{\frac{2}{3}}b\cat{0}+\sqrt{\frac{1}{3}}c\cat{-1}\Big)\otimes\Big(\ecat{1/2}+k\ecat{-1/2}\Big)
\]
in the case of satisfaction of equations
\[a=k^3d;\ b=\sqrt{3}k^2d;\ c=\sqrt{3}kd;\ d=\left(1+\abs{k}^2\right)^{-3/2}.
\]
Set of such states is topologically equivalent to sphere, and opposite in diameter points of the sphere correspond to orthogonal states forming the basis of not entangled states of four-dimensional irreducible representation.

Completely disentangled basis can be build with use of vectors of both subspaces of irreducible representations only.

\section{Density matrix eigenbasis}
 
 Let us now consider basis consisting of eigen vectors of density matrix of composite channel
\begin{equation}
	\hat{\rho}\catt{k}=\rho_k\catt{k};\ 1\geq \rho_1\geq\ldots\geq\rho_N; \ 
	\sum_{k=1}^N{\rho_k}=1.
\end{equation}

Let us construct ladder operators $\hat{J}_a$ and irreducible representations in bases of subsystems for this very basis. In this basis density matrix is sum of projectors to states $\catt{j,m}$
\begin{equation}\label{diag}
	\hat{\rho}_{sys}=\sum_{j=l-s}^{l+s}{\sum_{m=-j}^{j}{
	\rho_{j,m}\proa{j,m}
	}}.
\end{equation}

Properties of quantum system essentially depend on residual degrees of freedom in the density matrix representation. 

In the case of all the non-zero eigenvalues being different this decomposition is practically one, and statistical properties of system, along with entanglement of subsystem states, is completely determined by non-zero eigenvalues of density matrix and Clebsch-Gordan coefficients for respective eigenvectors. 

Degeneration of non-zero eigenvalues of density matrix brings additional degrees of freedom since linear combination of eigenvectors with same eigenvalue is an eigenvector as well, and this does not have all the properties following from \eref{clebsh}.

 \subsection{Nondegenerated channels}
 
We begin classification of states of two-particle quantum information transfer channel from study of channels non-degenerated from standpoint of statistical properties. Let us suppose that all the nonzero eigenvalues of density matrix are different.

\subsubsection{Pure states}
The simpliest among nondegenerated states are pure states with density matrix with one non-zero eigenvalue only, $\rho_{j,m}=1$
\begin{equation}\label{p_diag}
	\hat{\rho}_{j,m}=\proa{j,m}=\sum_{k,n=-s\ldots s}{C_{j,m;k}C_{j,m;n}
	\cat{m-k}\bra{m-n}\otimes\ecat{k}\ebra{n}
	}
\end{equation}
Density matrices of particle states
\[
\hat{\rho}^{\left\{B\right\}}_{j,m}=\sum_{m_s=-s\ldots s}{C_{j,m;m_s}^2
\roa{m-m_s}};\] 
\[\hat{\rho}^{\left\{A\right\}}_{j,m}
=\sum_{m_s=-s\ldots s}{C_{j,m;m_s}^2
\eroa{m_s}}.
\]
are same, and can be given by matrix
\begin{equation}\label{sub_rho}
\hat{\rho}^{\left\{sub\right\}}_{j,m}=\sum_{n=-s\ldots s}{\rho_{j,m;n}
\roa{n}};\ \rho_{j,m;n}=C_{j,m;n}^2.
\end{equation}
Both matrices are diagonal since in each term $\cat{m-k}\otimes\ecat{k}$ 
of the sum \eref{clebsh} vectors of state of each particle are orthogonal to any other vector of that sum, and non-diagonal terms turn to zero at averaging state vector of composite quantum information transfer channel by states of each of particles of the channel separately. 	

State density matrix of particle  \eref{sub_rho} has range equal to dimension of the smaller subsystem.

Correlation properties are determined by non-diagonal terms \eref{p_diag}, i.e. terms of double sum with different indices $k$, $n$. 
Probabilities $P_{n}$ of registration of one of the particles in state $n$ is equal to average value of projector $\roa{n}$ or $\eroa{n}$ 
\begin{equation}
	P_{n}^{\left\{A\right\}}=C_{j,m;m-n}^2;\ P_{n}^{\left\{B\right\}}=C_{j,m;n}^2.
\end{equation}
Probability $P_{k,n}$ of simultaneous registration of one particle in state $k$ and the other particle in state $n$ is equal to average value of projector $\roa{k}\otimes\eroa{n}$
\begin{equation}
	P_{k,n}=C_{j,m;n}^2 \delta_{m-n,k}.
\end{equation}
Accordingly, conditional probability of particle ${\left\{A\right\}}$ detection in state $k$ in the case of particle ${\left\{B\right\}}$ being in state $n$ is equal to one if $n+k=m$ and zero in all the other cases. 

So, states of particles are completely correlated if two-particle quantum information transfer channel is in one of pure states \eref{diag}. there are just two independent pure states $\proa{j,\pm j}$, all the other pure states are entangled.

Correlation between particles in independent states is somewhat conventional since for each of particles probability of one value of moment projection is equal to one, and of all the others -- to zero, irrespectively of state of the other particle. 
 
\paragraph{Paraqubits}
Two entangled states of paraqubit give for particles same equally distributed mixed states with density matrices $\frac{1}{2}\hat{1}$.  
\paragraph{Paraqutrits} 
There are four entangled basis states (of six) Those have such one-particle density matrices:
\begin{equation}
	\begin{array}{lll}
	\catt{\frac{1}{2},\pm\frac{1}{2}}:& \frac{1}{3}\roa{0}+\frac{2}{3}\roa{\pm1};&
	\frac{1}{3}\eroa{\pm\frac{1}{2}}+\frac{2}{3}\eroa{\mp\frac{1}{2}}\\
	\catt{\frac{3}{2},\pm\frac{1}{2}}:& \frac{2}{3}\roa{0}+\frac{1}{3}\roa{\pm1};&
	\frac{2}{3}\eroa{\pm\frac{1}{2}}+\frac{1}{3}\eroa{\mp\frac{1}{2}}
	\end{array}
\end{equation}
Density matrix of states of each qutrit particles have only two nonzero eigenvalue each, and equal distribution of states is not present.
\subsubsection{Mixed states}
Mixed state is a weighted mix of pure states. In the basis of eigenstates density matrix is diagonal, and with regard to properties of symmetry of the basis it can be given by weighted sum of irreducible representations of group $U(2)$ density submatrices
\begin{equation}
	\hat{\rho}_{sys}=
	\sum_{j=l-s}^{l+s}{\sum_{m=-j}^{j}{p_{j,m}\hat{\rho}_{j,m}	}};\ 
	\sum_{j=l-s}^{l+s}{\sum_{m=-j}^{j}{p_{j,m}	}}=1
\end{equation}

\begin{equation}\label{full}
	\hat{\rho}_{sys}=\sum_{j=l-s}^{l+s}{\sum_{m=-j}^{j}{p_{j,m}\sum_{k,n=-s}^{ s}{C_{j,m;k}C_{j,m;n}
	\cat{m-k}\bra{m-n}\otimes\ecat{k}\ebra{n}}}
	}
\end{equation}

Density matrices of subsystems are diagonal since diagonal are density matrices of subsystems for each pure state.

\begin{equation}
	\hat{\rho}_{A}=
	\sum_{j=l-s}^{l+s}{\sum_{m=-j}^{j}{\sum_{n=-s}^{s}{p_{j,m}C_{j,m;n}^2
\roa{m-n}}	
	}}
\end{equation}

For the smaller subsystem through replacement of the order in sum calculation 
\begin{equation}
		\hat{\rho}_{B}=\sum_{n=-s}^{s}{\Big[\sum_{j=l-s}^{l+s}{\sum_{m=-j}^{ j}{p_{j,m}C_{j,m;n}^2\Big]
\eroa{n}}	
	}}
\end{equation}
one can easily obtain probabilities $p_n$ for registration of states of particle as coefficients at diagonal terms of density matrix
\begin{equation}
		p_n=\sum_{j=l-s}^{l+s}{\sum_{m=-j}^{ j}{p_{j,m}C_{j,m;n}^2}}.
\end{equation}
Similarly, probability of registration of states of the second particles are
\begin{equation}
		p_k=\sum_{j=l-s}^{l+s}{\sum_{m=-j}^{ j}{p_{j,m}C_{j,m;m-k}^2}}.
\end{equation}
Probability $P_{k,n}$ of simultaneous registration of one particle in state $k$ and the other one in state $n$ is equal to average value of projectors product $\roa{k}\otimes\eroa{n}$ and can be calculated by formula
\begin{equation} 
	P_{k,n}=\sum_{j=l-s}^{l+s}{p_{j,n+k}C_{j,n+k;n}^2 },
\end{equation}
where one should take into consideration that due to properties of Clebsch-Gordan coefficients, $C_{j,m>j;n}=0$ і $C_{j,m<-j;n}=0$.

Conditional probability of detection of particle  ${\left\{A\right\}}$in state $k$, in the case of particle ${\left\{B\right\}}$ being in state $n$, depends on probability distribution and differs from zero for all the $k$, $n$, for which $\sum_{j}{p_{j,k+n}}$ is above zero. At the same time this conditional probability is less than one if probabilities of states $p_{j,k+n}$ are above zero for several different values $j$, i.e. total correlation between particles takes place in specific cases like pure states only.

Main conclusion from analysis of two-particle quantum information transfer channel carried out is in exclusiveness of entangled states.

\section{Disentanglement of states}
Classification of states of two-particle quantum information transfer channel formulated above gives us possibilities to study effectively properties of real channels, with the  most important one -- entanglement of states -- among those.

Information transfer channels with classical properties are same as all the other quantum channels, though in the process of information transfer all the time (or almost all the time) those are in states that do not have quantum peculiarities, thus being not entangled. Quantum information transfer channel (at least two-particle one) has only two pure states that are not entangled, and it would be a too high evaluation of perfectness of information transfer engineering to hope that technique of channel state creation is so exact that there is no mixing of some part of entangled states to non-entangled ones; there must be other reasons for absence of entanglement in typical states of classical information transfer channels. 

\subsection{Degeneration of density matrix}

Arbitrary mixed states are entangled, that's why reason of absence of entanglement can be in the fact that density matrix as Hermitian matrix with eigenvalues and eigenvectors belongs to set of matrices with special properties only. To special properties of Hermitian matrix in finite-dimensional space only degeneration of eigenvalues relates.

Totally degenerated density matrix is proportional to unit matrix that is product of unit matrices of subsystems, that's why it is product of density matrices of subsystem states, thus corresponds to independent subchannels of information transfer. Degeneration is specific, besides that, to each pure state in which density matrix is $N-1$-times degenerated ($\rho_k^{deg}=0$), though among pure states only two correspond to independent information transfer subchannels. So, only for some pairs of eigenvalues result of density matrix degeneration is in disentanglement of state.  

\subsubsection{Paraqubit disentanglement}

The simplest example of two-particle quantum information transfer channel -- paraqubit -- is a rather transparent example of density matrix degeneration effect on entanglement of states.

General state of paraqubit has density matrix as a mix 
\begin{equation}\label{full_pq}
	\hat{\rho}=p_s	\hat{\rho}_s + p_{00}\proa{0,0}+ p_{11}\proa{1,1}+ p_{0}\proa{0}
\end{equation}
with part $p_s$ of singlet state, parts $p_{00}$ and $p_{11}$ of states of independent particles, and part $p_{0}$ of entangled triplet state
\begin{equation}
	p_s	=1- p_{00}- p_{11}- p_{0};\ p_{00}\geq0;\ p_{11}\geq0;\ p_{0}\geq0;\ p_s\geq0.
\end{equation}
In induced basis this very matrix takes the form
\begin{equation}\label{full_pqs}
\begin{array}{l}
	\hat{\rho}=	p_{00}\roa{0}\otimes\eroa{0}+p_{11}\roa{1}\otimes\eroa{1}\\
	 +	\frac{p_{0}+p_s}{2}\big(\roa{0}\otimes\eroa{1}+\roa{1}\otimes\eroa{0}\big)\\
+	\frac{p_{0}-p_s}{2}\big(
\cat{0}\bra{1}\otimes\ecat{1}\ebra{0}+
\cat{1}\bra{0}\otimes\ecat{0}\ebra{1}
\big)
\end{array},
\end{equation}
where one can easily see the part responsible for entanglement of states (the last line). This part vanishes in the case of coincidence of the part of singlet and the part of entangled triplet states, $p_s=p_{0}$.

Disappearance of entanglement of states in the example given takes place due to degeneration of density matrix -- coincidence of eigenvalues $p_s$ and $p_{0}$, corresponding to entangled states. Along with that, coincidence of other pairs of eigenvalues does not produce disentanglement of states.  

\subsubsection{Paraqutrit disentanglement}

Arbitrary state of paraqutrit is a weighted mix of pure states and can be given as
\begin{equation}
	\begin{array}{r}
	\hat{\rho}_{p}= \sum_{m=\pm1/2}{p_{1/2,m}\proa{1/2,m}}\\
	+\sum_{m=\pm1/2,\pm3/2}{p_{3/2,m}\proa{3/2,m}}
\end{array}
\end{equation}
In induced basis 
\begin{equation}
	\begin{array}{r}
	\hat{\rho}_{p}=
	 p_{3/2,+3/2}\roa{1}\otimes\eroa{1/2}+p_{3/2,-3/2}\roa{-1}\otimes\eroa{-1/2}\\
+\left(\frac{1}{3}p_{3/2,1/2}+\frac{2}{3}p_{1/2,1/2}\right)\roa{1}\otimes\eroa{-1/2}\\
+\left(\frac{1}{3}p_{3/2,-1/2}+\frac{2}{3}p_{1/2,-1/2}\right)\roa{-1}\otimes\eroa{1/2}\\
+\left(\frac{2}{3}p_{3/2,1/2}+\frac{1}{3}p_{1/2,1/2}\right)\roa{0}\otimes\eroa{1/2}\\
+\left(\frac{2}{3}p_{3/2,-1/2}+\frac{1}{3}p_{1/2,-1/2}\right)\roa{0}\otimes\eroa{-1/2}\\
+\frac{\sqrt{2}}{3}
\Big[
\left(p_{3/2,1/2}-p_{1/2,1/2}\right)
\left(\cat{1}\bra{0}\otimes\ecat{-1/2}\ebra{1/2}
+\cat{0}\bra{1}\otimes\ecat{1/2}\ebra{-1/2}\right)
\Big.\\
+\Big.
\left(p_{3/2,-1/2}-p_{1/2,-1/2}\right)
\left(\cat{-1}\bra{0}\otimes\ecat{1/2}\ebra{-1/2}
+\cat{0}\bra{-1}\otimes\ecat{-1/2}\ebra{1/2}\right)
\Big]
\end{array}
\end{equation}
in square brackets terms responsible for entanglement of states are combined. There is possible partial ($p_{3/2,\pm1/2}=p_{1/2,\pm1/2};\ p_{3/2,\mp1/2}\neq p_{1/2,\mp1/2}$) or total ($p_{3/2,1/2}=p_{1/2,1/2};\ p_{3/2,-1/2}= p_{1/2,-1/2}$) disentanglement of states.

Among partially degenerated states ones are distinguished in which contribution of two basis states only is left, for instance, state mix $\proa{3/2,3/2}$ and $\proa{3/2,1/2}$ with probabilities $(1\pm d)/2$.
\begin{equation}
	\begin{array}{r}
	\hat{\rho}_{p}=
+\frac{1}{2}\left(1-d/3)\right)\roa{1}\otimes\eroa{-1/2}
+\frac{1}{2}\left(1+d/3)\right)\roa{0}\otimes\eroa{1/2}\\
+\frac{\sqrt{2}}{3}d
\Big[
\cat{1}\bra{0}\otimes\ecat{-1/2}\ebra{1/2}
+\cat{0}\bra{1}\otimes\ecat{1/2}\ebra{-1/2}
\Big]
\end{array}
\end{equation}

\begin{figure}[h]
	\includegraphics[width=3in]{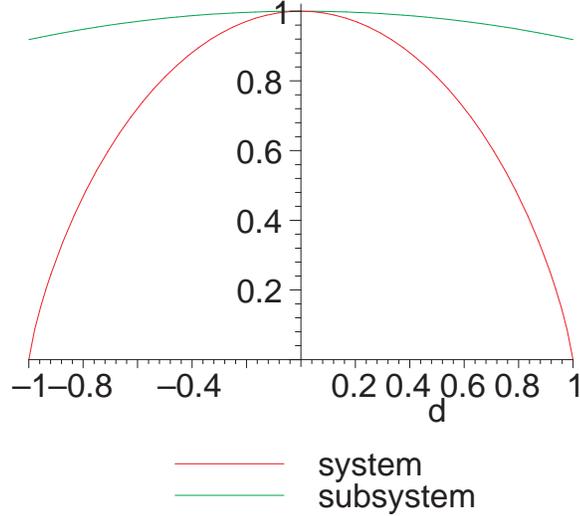}
	\caption{Dependence of entropies of paraqutrit and its qubit on non-degeneration of state. }
	\label{fig}
\end{figure}
 Entropy of such state and its subsystems is
\[S_{sys}=\sth{\frac{1+d}{2}},
\]
\[S_{A}=S_{B}=\sth{\frac{1+d/3}{2}}\]
Total degeneration of eigenvalues of density matrix corresponds to the value of non-degeneration measure $d=0$. Graph \ref{fig} shows dependence of entropy of paraqutrit and one of its particles on the measure of non-degeneration of state. To totally degenerated state same values of entropy of paraqutrit and one of its particles correspond, while arbitrary deflection from degeneration is accompanied by excess of qubit entropy over entropy of the system.

\section{Conclusions}

Two-particle parallel quantum information transfer channels can be classified by the level of degeneration of nonzero eigenvalues of density matrix. 

Among non-degenerated states two-dimensional set of not entangled states is distinguished, all the other states are entangled. 

Degeneration of eigenvalues in specific cases can be accompanied by loss of entanglement (disentanglement) of state.

Excess of entropy of each of subchannels over entropy of parallel quantum information transfer channel makes evidence of entanglement of states.

\paragraph{Acknowledgements}
Author would like to thank Organizing Committee of NATO Advanced Research Workshop "Quantum Communication and Security", Gdansk, POLAND, 10-13 September 2006, and personally its NATO Countries Director Janusz Kowalik, NATO Partner Countries Director Sergei Kilin and Chair of Program Committee Marek Zukowski for possibility for report and Pawel Horodecki for discussion.

\end{document}